# Large spin-charge interconversion induced by interfacial spin-orbit coupling in a highly conducting all-metallic system

Van Tuong Pham,[1,2] Haozhe Yang,[1] Won Young Choi,[1] Inge Groen,[1] Andrey Chuvilin,[1,3] F. Sebastian Bergeret,[4,5] Luis E. Hueso,[1,3] Ilya V. Tokatly,[3,5,6] and Fèlix Casanova [1,3,*]

[1]*CIC nanoGUNE, 20018 Donostia-San Sebastián, Basque Country, Spain*
[2]*University Grenoble Alpes, CEA, CNRS, Spintec, F-38000 Grenoble, France*
[3]*IKERBASQUE, Basque Foundation for Science, 48009 Bilbao, Basque Country, Spain*
[4]*Centro de Física de Materiales CFM-MPC (CSIC-UPV/EHU), 20018 Donostia-San Sebastian, Basque Country, Spain*
[5]*Donostia International Physics Center (DIPC), 20018 Donostia-San Sebastián, Basque Country, Spain*
[6]*Nano-Bio Spectroscopy Group, Departamento de Polímeros y Materiales Avanzados: Física, Química y Tecnología, Universidad del País Vasco (UPV/EHU), 20018 Donostia-San Sebastián, Basque Country, Spain*

[*]*e-mail: f.casanova@nanogune.eu*

## Abstract

Spin-charge interconversion in systems with spin-orbit coupling has provided a new route for the generation and detection of spin currents in functional devices for memory and logic such as spin-orbit torque switching in magnetic memories or magnetic state reading in spin-based logic. Disentangling the bulk (spin Hall effect) from the interfacial (inverse spin galvanic effect) contribution has been a common issue to properly quantify the spin-charge interconversion in these systems, being the case of Au paradigmatic. Here, we obtain a large spin-charge interconversion at a highly conducting Au/Cu interface which is experimentally shown to arise from the inverse spin galvanic effect of the interface and not from the spin Hall effect of bulk Au. We use two parameters independent of the microscopic details to properly quantify the spin-charge interconversion and the spin losses due to the interfacial spin-orbit coupling, providing an adequate benchmarking to compare with any spin-charge interconversion system. The good performance of this metallic interface, not based in Bi, opens the path to the use of much simpler light/heavy metal systems.

## I. INTRODUCTION

One of the fundamental ingredients in state-of-the-art spintronics is the generation and detection of a pure spin current, an opposite flow of up-spin and down-spin electrons that allows transferring angular momentum with minimal charge carriers, by exploiting the spin-orbit coupling (SOC) in a nonmagnetic system that leads to spin-charge current interconversions. Charge-to-spin (CS) conversion arises from spin Hall effect (SHE) in bulk materials,[1] and interfacial inverse spin galvanic effect, also known as Edelstein effect (EE), at Rashba interfaces[2,3] and surface states of topological insulators.[4] Spin-to-charge (SC) conversion takes place with the corresponding reciprocal effects. Whereas the use of SC conversion to read out the magnetic state in spin-based logics[5] has been recently demonstrated,[6] the CS conversion is being widely applied to switch magnetic elements via spin–orbit torque (SOT),[7,8] which is promising for a second generation of MRAM memories with unmatched switching speed and endurance.[9]

In spite of the strong applied interest, quantification of these conversions is a common source of controversies[7,8,10.] In particular, the disentanglement of bulk and interfacial contributions is a crucial aspect for a proper quantification of spin-charge current interconversions[11,12]. Furthermore, the different dimensionality hinders direct comparison of efficiencies associated to bulk and interfacial effects. For instance, interfacial SOC arising from inversion symmetry breaking can potentially have larger CS conversion rate per volume than bulk heavy metals.[2,4] Recently, the spin galvanic effect, or inverse Edelstein effect (IEE), leading to SC conversion has been reported on the nonmagnetic interface of the all-metallic,[3,13,14,15,16,17] metal/oxide,[18,19,20,21] and all-oxide interfaces.[22,23,24,25] In particular, the SC conversion in non-magnetic metallic interfaces has been so far based on Bi, with some controversy on the origin of the effect due to the predominance of SHE in Bi [26], or on metals which already show large SHE [27,28].

Au is the prototypical metal where Rashba splitting has been observed in a clean (111) surface by means of angle resolved photoemission spectroscopy (ARPES).[29,30] This observation, together with its very high conductivity, makes Au a promising candidate to decrease current densities in CS conversion-based devices. Indeed, large spin-charge interconversion has been reported in very thin Au films[31,32,33] and first-principles calculations suggest it is related to an interfacial effect.[34] This interfacial contribution might be at the origin of the large dispersion of reported spin Hall angles in Au.[10] In this work, we exploit the interfacial SOC in a Au/Cu interface to obtain a large SC and CS efficiency. The spin absorption technique in lateral spin valves (LSVs) is used to quantify the spin-charge current interconversion in this system. The experimental results show a negative sign of the conversion rate, incompatible with a SHE origin in the Au bulk. The lack of SC conversion signals in bare Cu and in the Cu/Au/Cu double interface system, as well as the double spin absorption observed in Cu/Au/Cu, allow us to unequivocally conclude the spin-charge current interconversion arises from the interfacial SOC at the Au/Cu interface. We analyze the data using generalized boundary conditions to extract the proper conversion efficiency parameters that allows us to directly compare with bulk systems. Our demonstration of a large spin-charge current interconversion in a highly conducting all-metallic system not based in Bi opens the path to the use of much simpler light/heavy metal interfaces and our quantification will avoid artificially large conversion efficiencies, providing an adequate benchmarking for energy-efficient spin-based devices.

## II. RESULTS AND DISCUSSION

Nonlocal spin transport measurements are performed in LSVs to realize spin-charge current interconversion at the Au/Cu interface. The scanning electron microcopy (SEM) image of a device is shown in Fig. 1a. Two LSVs are built by three ferromagnetic (NiFe) electrodes ($F_1$, $F_2$, $F_3$) connected by a nonmagnetic (Cu) channel. The left LSV is used as a reference for the spin absorption experiment. The right LSV has a cross pattern in its channel. The horizontal Cu wire ($x$-direction) acts as the nonmagnetic channel of the LSV. On top of the vertical Cu wire ($y$-direction), a 3-nm-thick Au wire is deposited *in situ* by shadow evaporation. The Au layer is chosen to be much thinner than its spin diffusion length (~100 nm)[35,36] in order to minimize the SHE contribution from the bulk Au (see Appendix 1 for experimental details).

Figure 1a illustrates the spin absorption technique, well-established to estimate the pure spin current supplied to the SC conversion wire,[17,19,36,37,38,39] with two configurations corresponding to the nonlocal measurements of the LSVs on the SEM image of device D1. A charge current ($I_{app}$) is applied through the $F_2$/Cu interface, inducing a spin accumulation in the Cu channel. This will create a pure spin current, where the majority-spin electrons diffuse away from (and minority-spin electrons diffuse towards) the $F_2$/Cu vicinity. The diffusing spins relax over the spin diffusion length ($\lambda_{Cu}$) of the Cu channel. In open circuit conditions, a voltmeter connected through the $F_1$/Cu (in the reference configuration, blue circuit) or $F_3$/Cu (in the spin absorption configuration, red circuit) interface will probe the spin accumulation. The obtained voltage ($V_{NL}$), which is proportional to the spin accumulation at the detecting electrode, normalized to the applied current ($I_{app}$) is defined as the nonlocal resistance ($R_{NL} = V_{NL}/I_{app}$). An external field is applied along the easy axis to control the reversal of the two magnetizations. The value of $R_{NL}$ changes sign when the magnetic configuration of injector and detector ferromagnets changes from parallel ($R_{NL}^{P}$) to antiparallel ($R_{NL}^{AP}$). The difference $2\Delta R_{NL} = R_{NL}^{P} - R_{NL}^{AP}$ allows us to obtain the spin signal by removing any baseline resistance coming from non-spin related effects. The typical $R_{NL}$ measurements of the spin absorption LSV and its reference are shown in Fig. 1b. Since the injected spin current is absorbed while crossing the vertical Au/Cu channel, which acts as a spin sink, the obtained spin signal $2\Delta R_{NL}^{abs}$ is smaller than the reference spin signal $2\Delta R_{NL}^{ref}$.

The spin current absorbed in the Au/Cu nanowire can then be converted to a transverse charge current. Figure 1c sketches the configuration of the SC conversion measurement. By applying a charge current ($I_{app}$) from $F_2$, an $x$-polarized spin current is created and reaches the vertical Au/Cu wire, where it is absorbed along $z$-direction and converted into a charge current along $y$-direction. This is detected as a voltage ($V_{SC}$) by a voltmeter probing along the Au/Cu wire under open circuit conditions. The SC resistance ($R_{SC} = V_{SC}/I_{app}$) is determined as a function of an in-plane magnetic field along the hard axis of $F_2$ (Fig. 1d). By reversing the magnetic field, the opposite $R_{SC}$ is obtained, since the NiFe magnetization is reversed as well as the spin polarization of the spin current. The difference of the two values of $R_{SC}$ at the saturated magnetizations in the loop is the SC signal, denoted as $2\Delta R_{SC}$, and allows to remove any background signal. See Supplementary Note 1 for an independent extraction of the field at which the magnetization of the NiFe wire saturates. The obtained value at 10 K is $2\Delta R_{SC} = 58 \pm 6$ μΩ. Note that our Au/Cu system is highly conducting, $\rho_{Au} = 5.7$ μΩcm and $\rho_{Cu} = 2.4$ μΩcm at 10 K (see Appendix 2 for details on the extraction of the two resistivities).

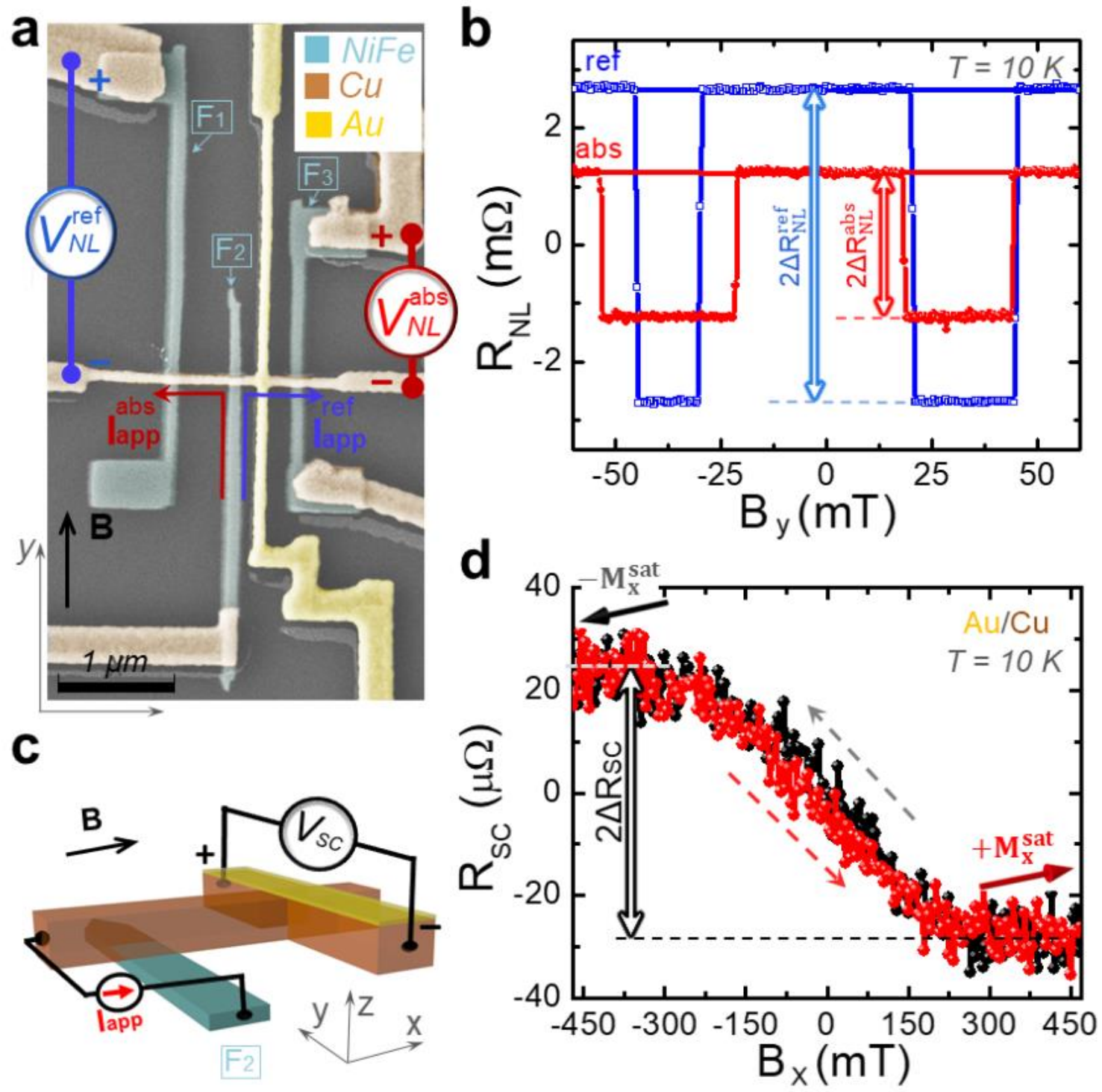

**FIG. 1**: a) **Spin absorption and SC conversion experiments.** SEM image of device D1 with two NiFe/Cu LSVs. The left one is a reference and the right one includes an Au/Cu wire between the two NiFe electrodes. The nonlocal measurement configurations (red and blue circuits correspond to the reference and spin absorption, respectively), the direction of the applied magnetic field ($B$) and the materials (NiFe, Cu and Au) are shown. b) Nonlocal resistance ($R_{NL}$) as a function of $B_y$ (trace and retrace) measured in device D1 at $I_{app}$= 200 μA and 10 K from the reference (blue squares) and the spin absorption (red spheres) configurations. From these measurements, we extract the spin signals $2\Delta R_{NL}^{ref}$ and $2\Delta R_{NL}^{abs}$. c) Sketch of the SC conversion measurement configuration. The direction of $B$ and the materials (NiFe, Cu and Au) are shown. d) Spin-to-charge resistance ($R_{SC}$) as a function of $B_x$ (trace and retrace) measured in device D1 at $I_{app}$= 200 μA and 10 K. Each curve is an average of 6 sweeps. From this measurement, we extract the SC signal ($2\Delta R_{SC}$).

A reciprocal configuration (see inset in Fig. 2a) allows us to measure the CS conversion, which is illustrated in Fig. 2a. The CS resistance loop shows an identical shape and same signal as the SC resistance loop in Fig. 1d, as expected from Onsager reciprocity.[40] The CS signal can be observed up to 300 K ($2\Delta R_{CS} = 8 \pm 5$ μΩ), as shown in Fig. 2b. To confirm the robustness of the spin-charge interconversion signals at the Au/Cu interfacial system, results were reproduced with three devices (D1, D2, and D3) in different substrates (see Fig. 3b).

Surprisingly, the sign of spin-charge interconversion signals is negative. A positive signal would be expected from the conventional SHE in bulk Au,[36,38] which consistently shows $\theta_{SH}$>0 (Ref. 10) or even from the positive skew scattering angle in Cu-Au alloy,[41] which could be forming at our interface. To confirm the negative sign of the signal, we carefully performed a control experiment with Pt as a spin absorption middle wire (see Appendix 3). This result rules out the bulk SHE as the main source of spin-charge interconversion, suggesting it is interfacial or superficial SOC of Rashba-type.

We next performed the control experiments to elucidate the origin of the SC and CS signals. The spin-charge current interconversion due to a naturally formed $CuO_x$/Cu interface in the bare part of Cu channel might be a source for the SC and CS signals.[42] For this reason, the CS conversion is measured on a device with an identical design (Fig. 1a), but where no Au is deposited, leaving the vertical channel also with bare Cu. The flat CS resistance, shown in Fig. 2c, indicates that the naturally oxidized Cu does not contribute to the measured SC and CS signals. Hence, the obtained signals must result from the influence of the 3 nm of Au on top of the vertical Cu wire.

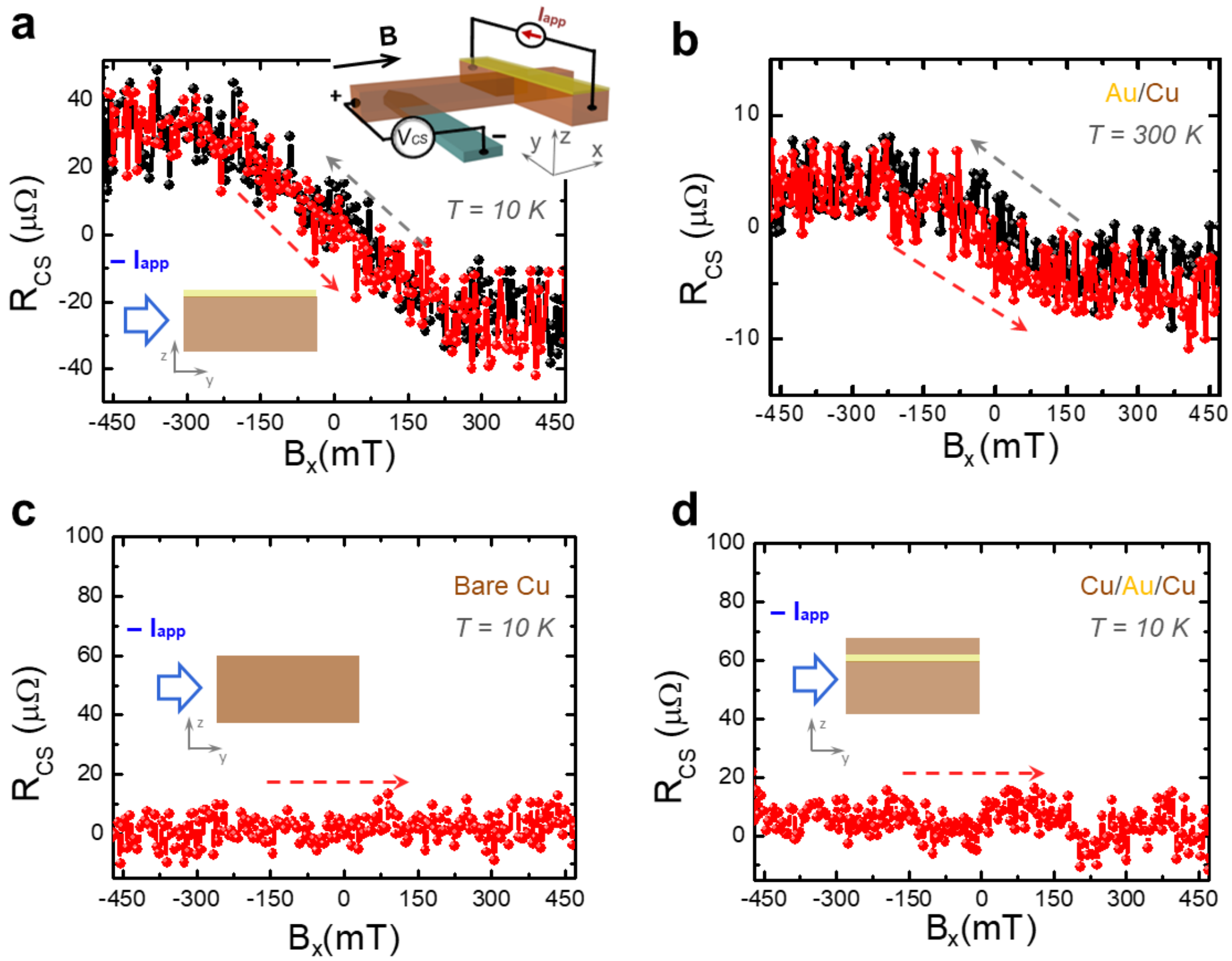

**FIG. 2: CS conversion and its interfacial SOC origin.** a) Charge-to-spin resistance ($R_{CS}$) as a function of $B_x$ (trace and retrace) measured in device D1 at $I_{app}$= 150 μA and 10 K. Each curve is an average of 3 sweeps. From this measurement, we extract the CS signal ($2\Delta R_{CS}$). Top-right inset: sketch of CS conversion measurement configuration. Bottom-left inset: sketch of the charge current flowing in plane at the Au/Cu interface. b) Same measurement as panel a, taken at 300 K. Each curve is an average of 8 sweeps. c) $R_{CS}$ as a function of $B_x$ measured at $I_{app}$= 150 μA and 10 K in a control device with bare Cu (no Au layer on top of the vertical Cu wire). Each curve is an average of 5 sweeps. Inset: sketch of the charge current flowing along the vertical Cu wire (in plane with the Cu surface). d) $R_{CS}$ as a function of $B_x$ measured at $I_{app}$= 150 μA and 10 K in a control device with a Cu/Au/Cu stack at the vertical wire. Each curve is an average of 5 sweeps. Inset: sketch of the charge current flowing in plane at the Cu/Au/Cu double interface.

Since SHE in Au bulk has been ruled out, the source of the spin-charge interconversion must be either the Au/Cu interface or the top Au surface. Interfacial SOC at a metallic interface/surface arises from the breaking of inversion symmetry giving rise to the inverse spin galvanic effect.[46] In order to confirm this point, we performed a control experiment with a device where a trilayer Cu/Au/Cu wire replaces the bilayer Au/Cu wire. The thickness of the thin Cu on top of Au is 7 nm. Since the stacking order reverses the orientation of the interface, the sign of the inverse spin galvanic effect is different at each interface provided the conversion is mediated by the interfacial SOC at Cu/Au interface. Therefore, it is expected that the conversion will vanish in a symmetric Cu/Au/Cu structure.[13,15] Alternatively, in a scenario where the conversion takes place at the Au/vacuum surface, capping the top Au surface with a Cu layer will suppress the effect by destroying the Au/vacuum Rashba surface. Both scenarios are compatible with the flat SC resistance shown in Fig. 2d. From this experiment, thus, we cannot distinguish the two possibilities.

The exact location of the spin-charge interconversion can be determined by the results of the spin absorption measurements. Figure 3a shows the spin absorption ratio ($\eta = \Delta R_{NL}^{abs}/\Delta R_{NL}^{ref}$) as a function of temperature measured in the 3 different systems (bare Cu, Au/Cu, and Cu/Au/Cu). $\eta$ is independent on the temperature suggesting that Elliott-Yafet is the dominating spin relaxation mechanism.[19,37] In the bare Cu system, $\eta_{Cu} = 0.96 \pm 0.07$ indicates that the spin absorption in the vertical Cu wire is negligible. In contrast, the spin absorption in the Au/Cu system is considerable, $\eta_{Au/Cu} = 0.43 \pm 0.06$, in agreement with the presence of spin-charge interconversion. Interestingly, the spin absorption is two times larger in the Cu/Au/Cu system ($\eta_{Cu/Au/Cu} = 0.22 \pm 0.06$), concluding that the spin current is absorbed mainly at the Au/Cu interfaces and not at the top Au surface.[43] While spin loss at surface or interface may have several origins (including spin-flip scattering) and may not be necessarily due to spin-charge interconversion,[44] spin-charge interconversion by the interfacial SOC necessarily involves spin loss[45], thus ~~we rule~~ruling out spin-charge interconversion at the Au surface. The absence of a net spin-charge interconversion in the Cu/Au/Cu system can thus be naturally explained by the inverse spin galvanic effect of opposite sign at the two interfaces with opposite orientation.

The temperature dependence of the SC signal is shown in Fig. 3b. The negative value is obtained at all measured temperatures, with a strong decay from 10 to 300 K. This is in contrast with the temperature independence of the spin absorption, although it is not surprising, since spin loss has different sources, as mentioned above. The spin absorption (Fig. 3a) allows us determining the spin current, whereas the SC signal (Fig. 3b) reveals how much it is converted to charge current. Based on these two separate measurements, we can estimate the spin loss and the efficiency of the SC conversion. We perform FEM simulations based on a two-current drift diffusion model (see Appendix 4 for details). The interface is modeled as a layer with a certain thickness $t_{int}$ on the Au side next to the Cu. The spin loss and the SC conversion efficiency are quantified by an effective spin diffusion length ($\lambda_{int}$) and an effective spin Hall angle, $\theta_{SH}^{eff}$, respectively. A strong temperature dependence of $\theta_{SH}^{eff}$ is obtained (Fig. 3c), following the same trend as the SC signal. The values of $\theta_{SH}^{eff}$ shown in Fig. 3c are calculated considering that $t_{int}$= 3 nm, *i.e.*, the full Au wire thickness effectively represents the interface. With this assumption, $\theta_{SH}^{eff}$ would be close to the ones reported for Pt using the same technique[36,37,39], but with opposite sign. However, the $\theta_{SH}^{eff}$ value depends on the choice of $t_{int}$ (see Table 1 in Appendix 4), evidencing that this effective value is not a proper quantity to characterize the spin-charge interconversion in this system. Note that similarly large values of the spin Hall angle have been reported in thin Au (Refs. 31, 32, and 33) but again with opposite sign. Therefore, any possible interfacial SCC contribution that enhances the bulk SHE of Au should have opposite sign to our Cu/Au interface.

For the spin-orbit applications[7], the large values of $\theta_{SH}^{eff}$ in a highly conducting system are expected to be beneficial. In order to confirm this, harmonic Hall voltage experiments have been performed in Py/Cu/Au and Py/Cu/Pt Hall bars (see Supplementary Note 2). Indeed, our results show that the Cu/Au system gives higher damping-like torque, which is the primary source for magnetization reversal, than Pt with identical geometrical parameters.

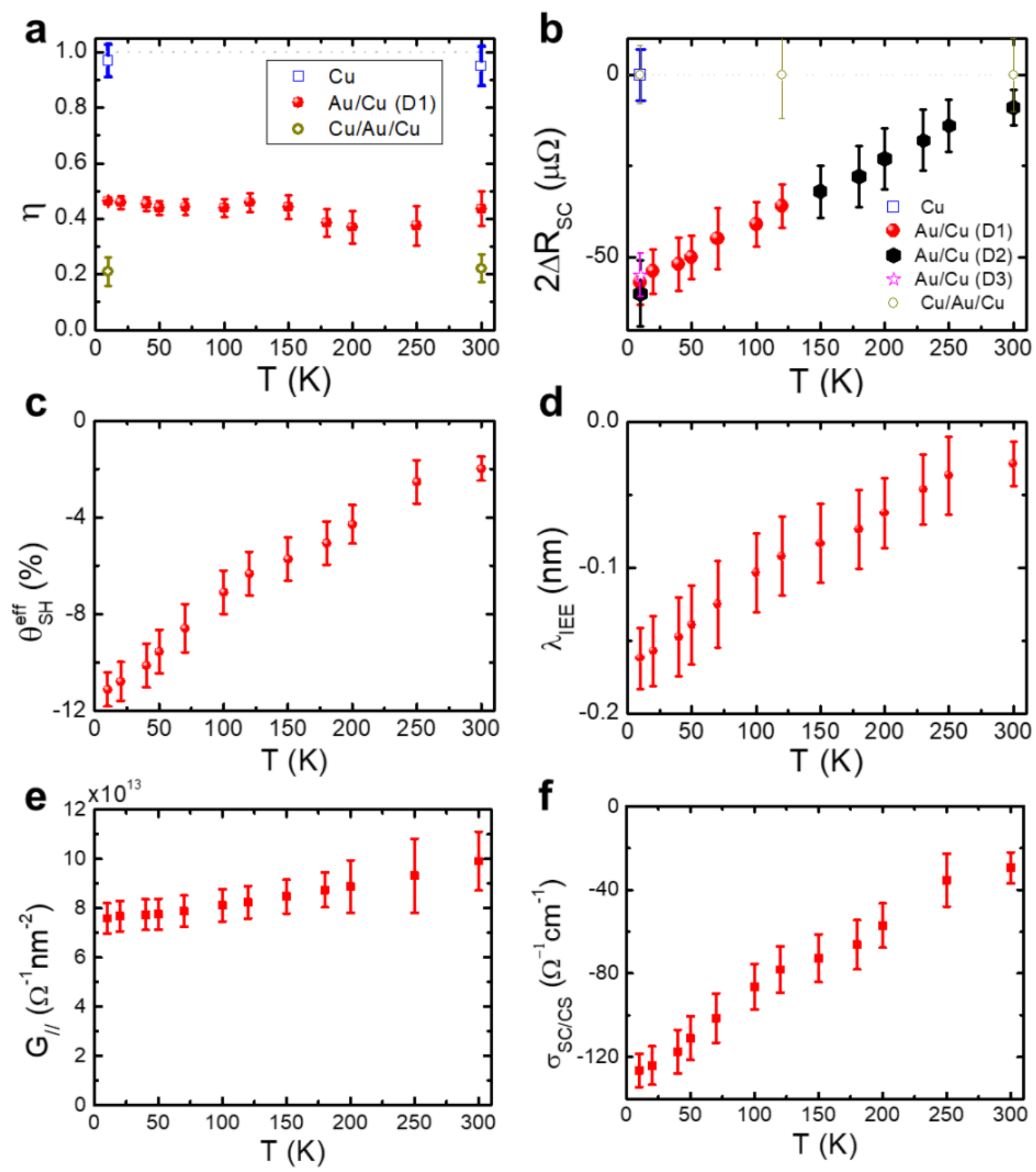

**FIG. 3**: **Temperature dependence of the spin absorption and SC conversion**. a) Spin absorption ratio ($\eta$) as a function of temperature obtained from the measured nonlocal signals for different devices. b) SC signal as a function of temperature for different devices. The negative SC signal at the Au/Cu system is reproduced in three different devices (D1, D2, D3) fabricated in two different substrates. c) Effective spin Hall angle ($\theta_{SH}^{eff}$) of our Au/Cu system as a function of temperature, obtained from data in panels a and b and modelling the interface by FEM with an effective thickness $t_{int}$= 3 nm. d) Inverse Edelstein length ($\lambda_{IEE}$), e) interfacial spin-loss conductance ($G_{\parallel}$), and f) interfacial spin-to-charge/charge-to-spin conductivity ($\sigma_{sc/cs}$) of our Au/Cu interface as function of temperature, extracted using the relations between the FEM and the boundary conditions models (see discussion above Eq. 3).

A more adequate quantification can be performed by assuming our SC conversion arises from an interfacial effect. This can be described by the effective boundary conditions derived in Ref. 46. Here, we adapt the simplest form of these boundary conditions inspired by the ones for metal/insulator interface.[19] Assuming that Au/Cu interface is located at $z = 0$, the boundary conditions read

$$J_z^x|_{0^-} - J_z^x|_{0^+} = G_\parallel \mu_s^x\big|_0 + \sigma_{cs}\partial_y \mu_c\big|_0 \tag{1}$$

$$j_z|_{0^-} - j_z|_{0^+} = \sigma_{sc}\partial_y \mu_s^x\big|_0 = \nabla \cdot \vec{j}_{int} \tag{2}$$

Here $\mu_c$ and $\mu_s^x$ are the charge and spin electrochemical potentials. These two equations describe the behavior of the currents at the interface. In Eq. (1), the discontinuity of the spin current $J_z^x$ is proportional to the charge-to-spin conversion quantified by the parameter $\sigma_{cs}$. In addition, the SOC also leads to spin losses at the interface described by the parameter $G_\parallel$.[45] The second equation describes the conservation of charge current $j_z$: the discontinuity of $j_z$ at the interface equals to the divergence of an interfacial current defined as $\vec{j}_{int} = -e\sigma_{sc}(\hat{z} \times \vec{\mu}_s)$. Onsager reciprocity mandates that $\sigma_{sc} = \sigma_{cs}$. As shown in Ref. 19, there is a connection between the interfacial parameters and those of the FEM simulation. Namely, $G_\parallel = \sigma_{Au} t_{int}/\lambda_{int}^2$ and $\sigma_{sc} = \theta_{SH}^{eff}\sigma_{Au} t_{int}^2/2\lambda_{int}^2$. The SC conversion efficiency, usually denoted as inverse Edelstein length ($\lambda_{IEE}$), is then given by

$$\lambda_{IEE} = \frac{\sigma_{sc}}{G_\parallel} = \frac{1}{2}\theta_{SH}^{eff} t_{int}. \tag{3}$$

The temperature dependence of $\lambda_{IEE}$, $\sigma_{sc}$ and $G_\parallel$ are shown in Figs. 3d, 3e, and 3f, respectively. Importantly, whereas $\theta_{SH}^{eff}$ strongly depends on the chosen value of $t_{int}$, the obtained $\lambda_{IEE}$, $\sigma_{sc}$ and $G_\parallel$ are essentially independent of the choice of $t_{int}$ (see Table 1 in Appendix 4 for details), confirming they are robust quantities to model our SC conversion. A proper comparison of the SC efficiency is $\lambda_{IEE}$ of an interface with $\theta_{SH}\lambda_s$ of a bulk system.[47] The value of $\lambda_{IEE}$ in Au/Cu at 10 K (–0.17 nm) is comparable to the ones reported in Cu/BiO$_x$[19] and Ag/Bi,[3] and to $\theta_{SH}\lambda_s$ of Pt.[37,48] When working with spin-orbit torques, the parameter that is usually considered relevant in a bulk system is the spin Hall conductivity $\sigma_{SH} = \theta_{SH}\sigma_{xx}$, since a high longitudinal conductivity $\sigma_{xx}$ lowers the power consumption.[48,49] The corresponding interfacial parameter is $\sigma_{cs}$, which in Au/Cu at 10 K [–126 (ħ/e)$\Omega^{-1}$cm$^{-1}$] is 3 times larger than in Cu/BiO$_x$,[19] but one order smaller than $\sigma_{SH}$ in the best systems, such as Pt[37,48] or Ta.[50] Both $\sigma_{sc}$ and $\lambda_{IEE}$ quickly decay with increasing temperature. Here, we should emphasize that when spin-charge interconversion occurs at an interface, it has to be quantified by the proper interface parameters ($\sigma_{sc}$ and $G_\parallel$, or its ratio $\lambda_{IEE}$), not with a spin Hall angle, a parameter valid for the bulk SHE, in order to avoid obtaining unphysical values ($\theta_{SH}^{eff} > 1$) associated to the particular choice of the effective thickness.

There are different mechanisms that could lead to the spin galvanic effect at nonmagnetic interfaces. One possibility is the Rashba splitting at the interface band.[51] Assuming such origin of the effect, we can calculate the Rashba coefficient $\alpha_R$ from the obtained $\lambda_{IEE}$ using the relation $\alpha_R = \hbar\lambda_{IEE}/\tau$,[52] where the Au momentum scattering time ($\tau$) is extracted from its resistivity (Appendix 2). We obtain the effective values for our Au/Cu interface $\alpha_R = 0.52$ eVÅ and $0.22$ eVÅ at 10 K and 300 K, respectively. Another possibility is the spin-dependent scattering of the Bloch bulk states from the interface, sometimes called spin-orbit filtering.[53,54,55,56,57] Although we cannot distinguish experimentally between different microscopic mechanisms, we can explain the observed interfacial spin galvanic effect by the boundary conditions (Eqs. 1 and 2) and, independently of the origin, quantify it by two phenomenological parameters: the interfacial spin-to-charge conductivity $\sigma_{sc/cs}$ and the interfacial spin-loss conductance $G_{\parallel}$. A related effect is the interfacial SHE, which also originates from the interfacial SOC, as in the case of the Fe/Au interface.[34] However, the CS conversion studied in this work corresponds to a different experimental situation and therefore cannot be used to explain our observations. Particularly, the interfacial SHE is sensitive to the out-of-plane charge current whereas in our experiment an in-plane charge current is applied as shown in the inset of Fig. 2a.

## III. CONCLUSION

To conclude, we identified a sizable spin-charge interconversion at Au/Cu interfaces unequivocally arising from the interfacial SOC and not from bulk SHE in Au. The quantification of the results is based on effective boundary conditions with two parameters describing the spin-charge interconversion and the spin losses due to the interfacial SOC. The form of these boundary conditions is independent of the microscopic details, providing an adequate benchmarking to compare with any spin-charge conversion system. The inverse Edelstein length ($\lambda_{IEE} = -0.17$ nm) of Au/Cu interface is comparable to other all-metallic and metal/oxide interfaces, although it is the first one reported that is not based on Bi, which is an advantage for the fabrication process. Our finding of efficient spin-charge current interconversion in a highly conducting system with simple *3d* metals invigorates research towards the development of energy-efficient spin-based devices such as SOT-MRAMs[9] and spin-based logics[58].

## Acknowledgments

The authors would like to thank Dr. Alain Marty, Dr. Laurent Vila, Dr. Juan Borge, Dr. Ka Shen, and Prof. Ke Xia for fruitful discussions. This work is supported by Intel Corporation through the Semiconductor Research Corporation under MSR-INTEL TASK 2017-IN-2744 and the 'FEINMAN' Intel Science Technology Center, and by the Spanish MICINN under the Maria de Maeztu Units of Excellence Programme (MDM-2016-0618) and under project numbers RTI2018-094861-B-100 and FIS2017-82804-P. I.V.T. acknowledges support by Grupos Consolidados UPV/EHU del Gobierno Vasco (Grant No. IT1249-19). V.T.P. acknowledges postdoctoral fellowship support from 'Juan de la Cierva—Formación' programme by the Spanish MICINN (grant number FJCI-2017-34494).

## Appendix 1: Nanofabrication, material characterization and transport measurements

In this appendix, we describe the nanofabrication of the devices, the characterization of the materials used in these devices and the transport measurements performed.

The LSV devices used in this work (see FIG.4a) were fabricated with two-step e-beam lithography, metal deposition, and lift-off process on thermally oxidized $SiO_2$/Si substrates. Three ferromagnetic electrodes (F1, F2, F3) made of permalloy ($Ni_{80}Fe_{20}$), 30-nm thick and 90-nm wide, are fabricated in the first step. Nucleation pads are designed in F1 and F3 to favor a lower coercive field than the one in F2, with sharp ends. In the second step, the Cu cross is defined with a width of 80 nm, with the horizontal wire (along the *x* direction) connecting the 3 ferromagnetic electrodes, and 80 nm of Cu are deposited normal to the sample surface. Subsequently, the sample is tilted 45$^{o}$ from the normal in the *y-z* plane and 3 nm of Au are deposited. The shadowing effect allows us to cover only the vertical wire of the Cu cross along the *y* direction, whereas the horizontal Cu wire remains free from Au, as shown in FIG.4b. Figure 4c shows the sketch of the shadowing effect protecting the horizontal Cu wire under the 45° deposition of Au. Figure 4d is the SEM image of a wider region in the horizontal Cu channel. The 45° deposition can hide a 236-nm-wide nanowire of Cu under the shadow of ZEP resist. Since the width of the horizontal Cu channel at the relevant LSV region is only 80 nm, it is not covered by the deposited Au.

NiFe is e-beam evaporated at 0.6 Å/s and $1.0\times10^{-8}$ mbar. Cu is thermally evaporated at 1.8 Å/s and $2.0\times10^{-8}$ mbar. Au is e-beam evaporated at 0.1 Å/s and $5.0\times10^{-9}$ mbar. Before the deposition of Cu, an Ar-ion milling process (Ar-ion flow with normal incidence, 15 sccm, acceleration voltage of 50 V, beam current of 50 mA, and beam voltage of 300 V) is performed for ~30 s to ensure transparent NiFe/Cu interfaces.

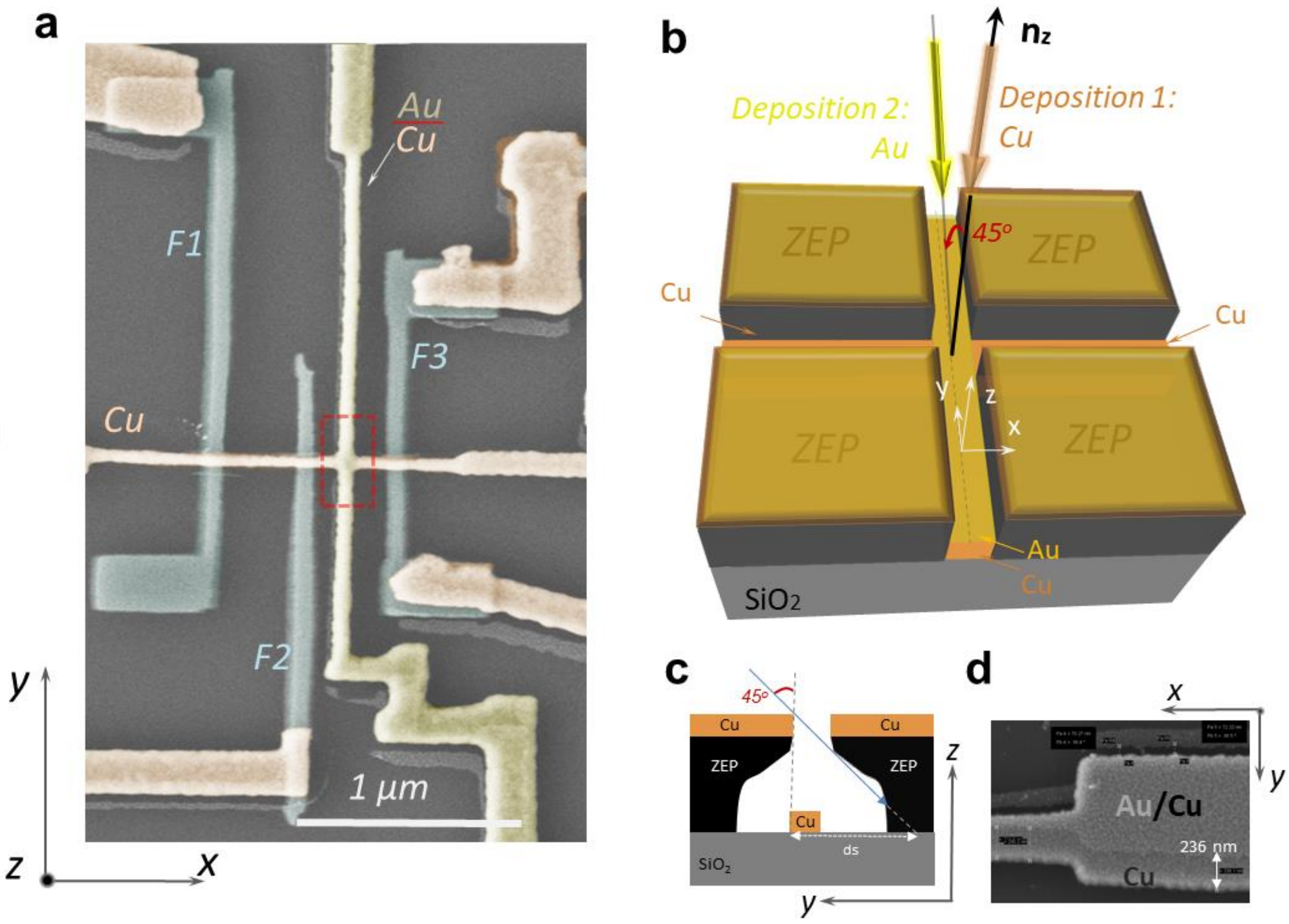

**FIG. 4:** **a)** False-colored SEM image of device D1 after finishing the nanofabrication. **b)** Sketch of the Cu and Au depositions performed at different angles, corresponding to the dashed area in panel a, which shows how Au is deposited in the vertical wire of the Cu cross only. **c)** *y–z* plane cut of the sketch in panel b, showing the ZEP resist profile and the horizontal Cu wire. Blue arrow shows the direction of Au deposition. $d_s$ is the shadow length that indicates the maximum width of the horizontal Cu wire that can be hidden by the shadow under the 45°-angle deposition. **d**) SEM image showing that the angle deposition works well in the nanodevice. The 80-nm-wide horizontal Cu wire is not covered by the deposited Au because $d_s$ = 236 nm in the experiment.

The structure of Au/Cu bilayer has been characterized by (Scanning) Transmission Electron Microscopy (STEM), Energy dispersive X-ray Analysis (EDX) and grazing-incidence X-ray diffraction. The results are presented in FIG.5.

Figure 5a shows a cross-sectional high-angle annular dark-field (HAADF) STEM image of the bilayer overlayed by an EDX map of the corresponding region. It clearly reveals a continuous, yet wavy, 3-nm-thick layer of Au on top of 80-nm-thick Cu layer. Grazing incidence X-ray diffraction (FIG.5b) reveals 3 main peaks: the most intensive (111) reflection of Cu, substantially weaker (002) peak of Cu, and (111) peak of Au. This clearly indicates a dominating (111) texture of the Cu layer and collinearly overgrown Au. High-resolution TEM image on FIG.5c confirms an epitaxial overgrowth of Au over Cu. The map of the (111) lattice parameter obtained from FIG.5c and shown in FIG.5d (Ref. [59]) reveals an expected expansion of the lattice when changing from Cu to Au.

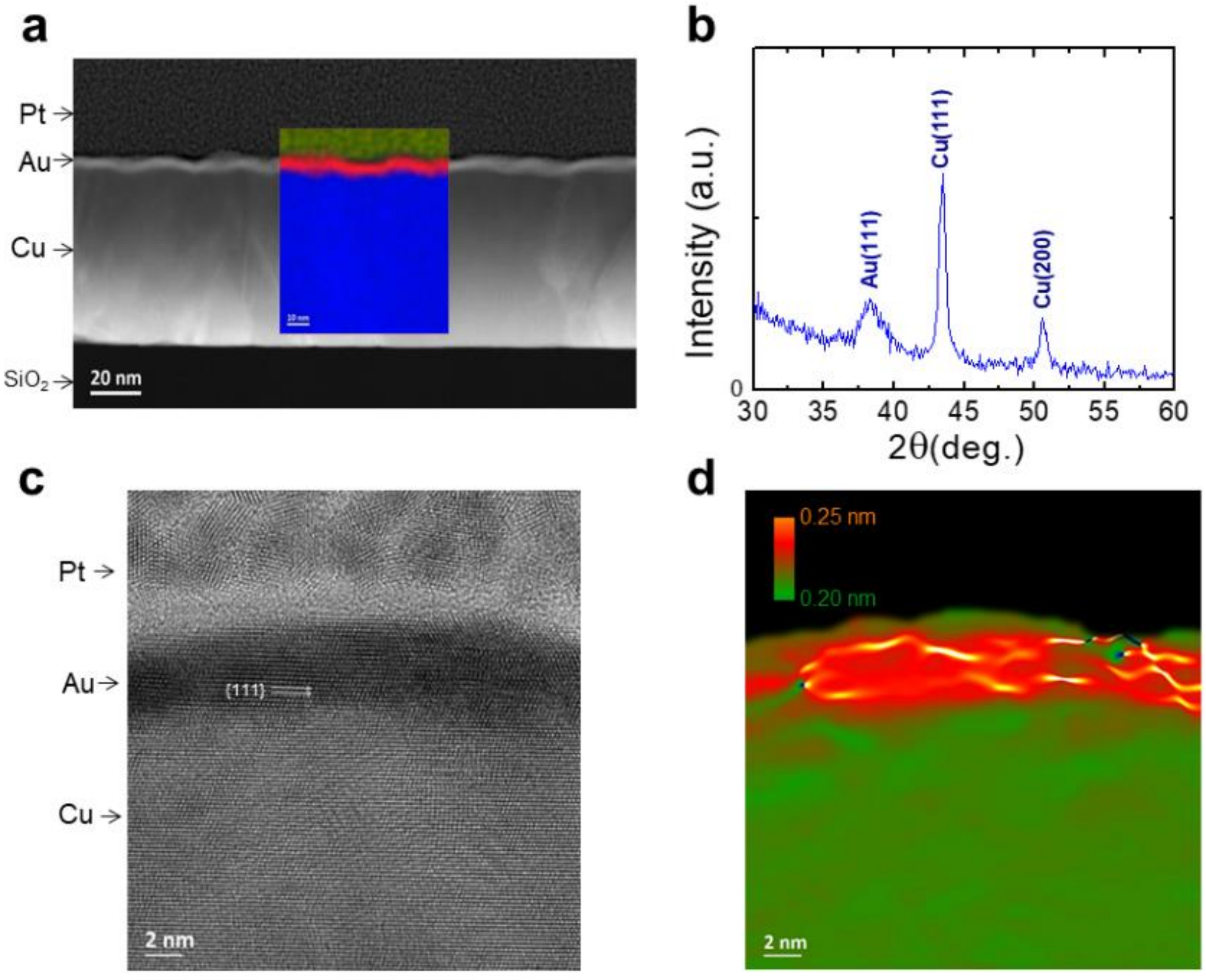

**FIG. 5: a)** Cross-sectional HAADF STEM image of Au(3)/Cu(80) bilayer and EDX map in the inset shows layers of Au (red), Cu (blue), and Pt protective layer (greenish). **b)** Grazing incidence X-ray diffraction with 2Θ scan for a grazing incidence angle of $\phi$=0.5° in a reference Au(3)/Cu(80) bilayer thin film. **c**) High resolution TEM image. **d**) Map of the (111) lattice parameter obtained from panel c.

Electronic transport measurements are performed in a Physical Property Measurement System from Quantum Design, using a 'DC reversal' technique with a Keithley 2182 nanovoltmeter and a 6221 current source at temperatures ranging from 10 to 300 K. The applied current $I_c^{app}$ for the nonlocal measurements is between 150 and 200 μA. We apply in-plane and out-of-plane magnetic fields with a superconducting solenoid magnet by rotating the sample using a rotatable sample stage.

## Appendix 2: Measurements of the Cu and Au resistivities

The resistivity of 3-nm-thick Au is obtained by comparing the resistances of Cu(10)/Au(3) and Cu(10)/ $SiO_2$(5) thin films, where the numbers stand for the thickness in nm. For this purpose, Hall bar structures are fabricated on two different Si substrates with thermally oxidized $SiO_2$(150) using a single step electron-beam lithography process. For the Cu(10)/$SiO_2$(5) sample, Cu is thermally evaporated in the same ultra-high vacuum system as the other samples used in this work. Subsequently, the Cu is capped by *ex-situ* (but very fast sample transfer between two chambers) sputtering $SiO_2$(5) to prevent gradual oxidation of the Cu in air. In the case of the Cu(10)/Au(3) sample, the same Cu evaporated is used, followed by *in-situ* deposition of Au via e-beam evaporation. In both samples, 1 nm of Ti is e-beam evaporated *in situ* as a seed layer to ensure a good adhesion of Cu to the substrate. This layer affects the two samples equally, so that there is no significant influence on the estimation of the Au resistivity.

We perform 4-probe electrical measurements to extract the resistance of the double Hall bars. Assuming that the Cu(10)/Au(3) film can be considered as two parallel connected resistances, the resistance of the Au(3) channel is given by Ohm's law:

$$R_{\mathrm{Au(3)}} = \left[\frac{1}{R_{\mathrm{Cu(10)Au(3)}}} - \frac{1}{R_{\mathrm{Cu(10)}}}\right]^{-1}$$

where $R_{\mathrm{Cu(10)Au(3)}}$ and $R_{\mathrm{Cu(10)}}$ are the resistances of the Cu(10)/Au(3) and Cu(10)/$SiO_2$(5) stacks, respectively.

Figure S4a shows the measured resistances of the Cu(10)/Au(3) (red) and Cu(10)/$SiO_2$(5) (blue) stacks and the calculated resistance of Au(3) (green). By considering the width ($w$) and the length ($L$) of the Hall bars, namely $w \sim 4100$ nm and the $L \sim 29400$ nm, and the proper thickness, the electrical resistivity is obtained from $\rho = R\,w\,t/L$. Figure S4b displays the resistivity for Cu(10) (blue) and Au(3) (green).

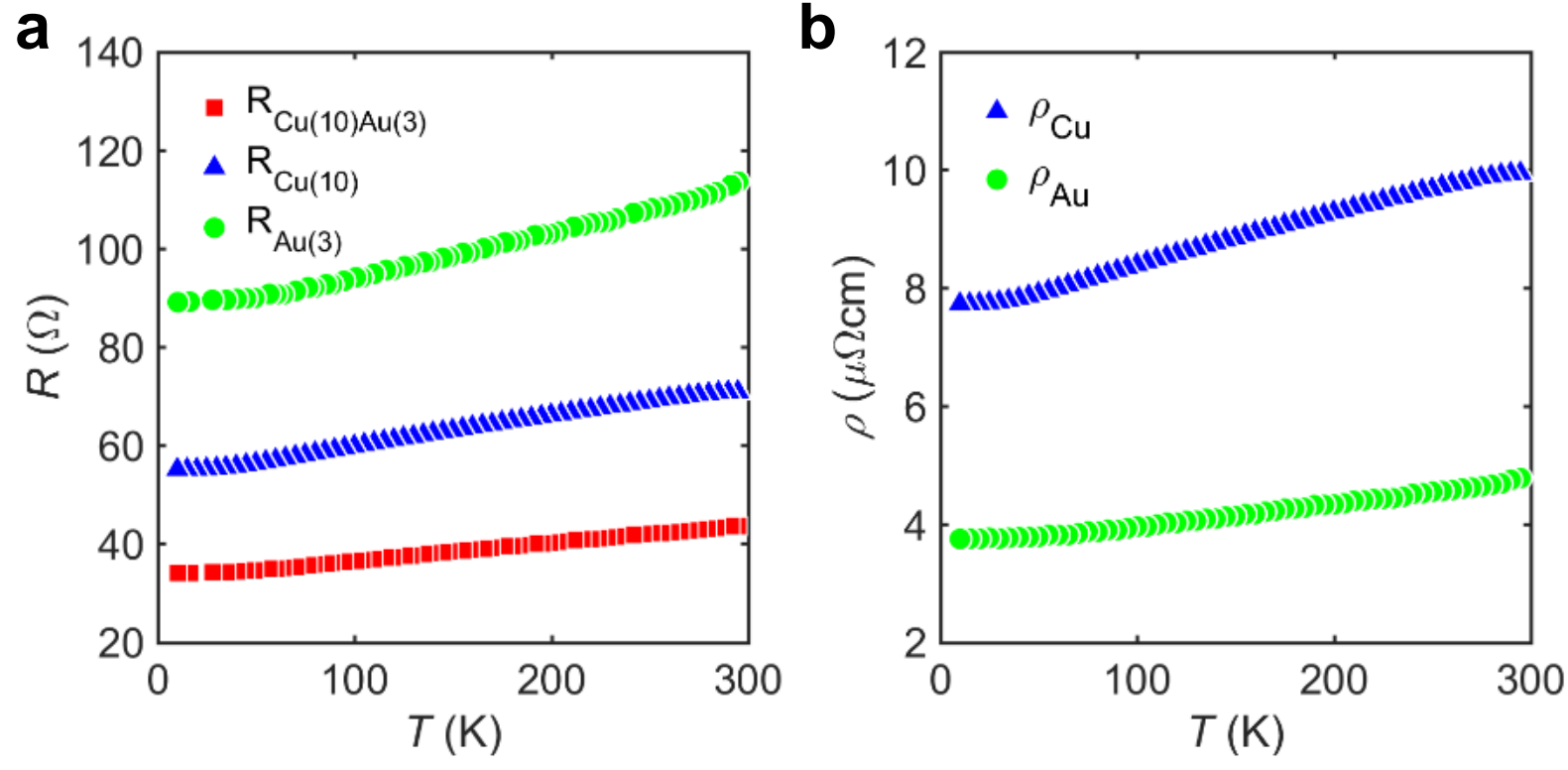

**FIG.6: a**) Resistance of Cu(10 nm)/Au(3 nm) (red) and Cu(10 nm)/$SiO_2$(5 nm) (blue) stacks shaped as double-Hall crosses with an identical lateral geometry, measured using the 4-probe measurement. The extracted value for the resistance of 3-nm-thick Au is shown in green. **b**) The calculated value of resistivity of the 3-nm-thick Au and 10-nm-thick Cu.

The resistivity of Cu channel can be measured by the 4-probe technique directly in the LSV device and the result as a function of temperature is plotted in FIG.7. The lower resistivity of Cu compared to that in FIG.6b is expected from the thicker Cu wire in the LSV.

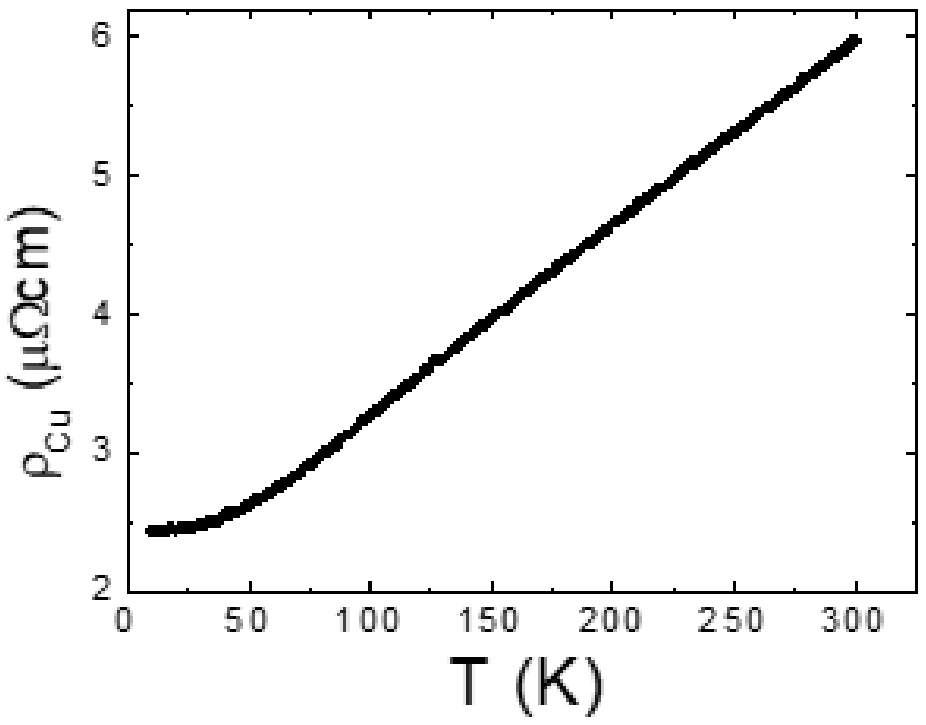

**FIG.7:** Resistivity of Cu channel of the LSV as a function of temperature.

## Appendix 3: Confirmation of the negative sign of spin-to-charge signal in Cu/Au

In order to confirm the negative sign of the SC signals in the Cu/Au wire, we performed a control experiment of the SC conversion in a Pt wire, which is well established to have a positive spin Hall angle.[60] A spin sink of 100-nm-wide Pt wire is fabricated under a single Cu channel, which replaces the Au on top of the transverse Cu wire as presented in the main text and Appendix 1. The SEM image of the sample is shown in FIG.7a.

Figure S6b compares the spin-to-charge resistance ($R_{SC}$) measured on a LSV with Au/Cu (device D1) and Pt (device D4) wires, using the same electrical measurement configuration. Note that, although the effective spin Hall angle for the two devices is quite similar, the SC signal ($2\Delta R_{SC}$) measured in the Pt wire is an order of magnitude higher than that of Cu/Au interface due to the shunting effect. Most importantly, since the two devices have opposite stacking order (Au is on top of the Cu spin channel in device D1, whereas Pt is below the Cu spin channel in device D4), the same sign of the SC signals produced demonstrates an opposite sign of the effective spin Hall angle.

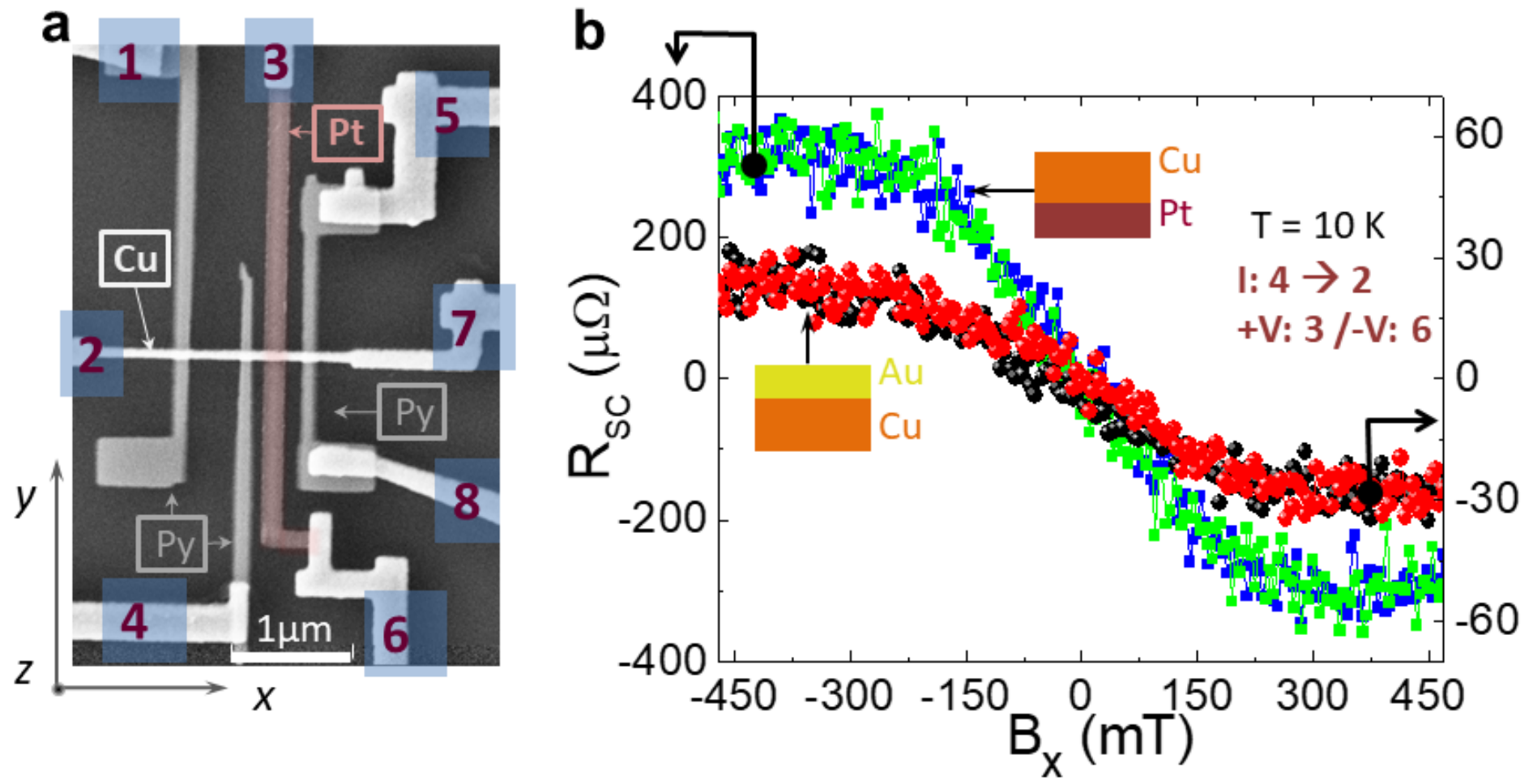

**FIG.8: a**) SEM image of the control device D4 showing the labeling of contacts used for the transport measurements. The device consists of two NiFe/Cu LSVs: the left one is a reference and the right one includes a Pt wire between the two NiFe electrodes. **b**) Spin-to-charge resistance ($R_{SC}$) as a function of $B_x$ (trace and retrace) measured at $I_{app}$= 200 µA and 10 K for the Au/Cu (device D1) and Cu/Pt (device D4) systems. Each curve is an average of 3 sweeps for device D1. The same sign of the SC signals is obtained with the same measurement configuration and the opposite stacking order.

**Appendix 4: 3D finite elements method simulations**

3D finite elements method (FEM) simulations are performed based on the two-current drift-diffusion model mostly using the formalisms for LSVs (see Ref. [61]). Figure S7a shows the geometry of the simulated device and the mesh of the finite elements. The geometry construction and 3D-mesh were elaborated using the free software GMSH [62] with the associated solver GETDP [63] for calculations, post-processing and data flow controlling.

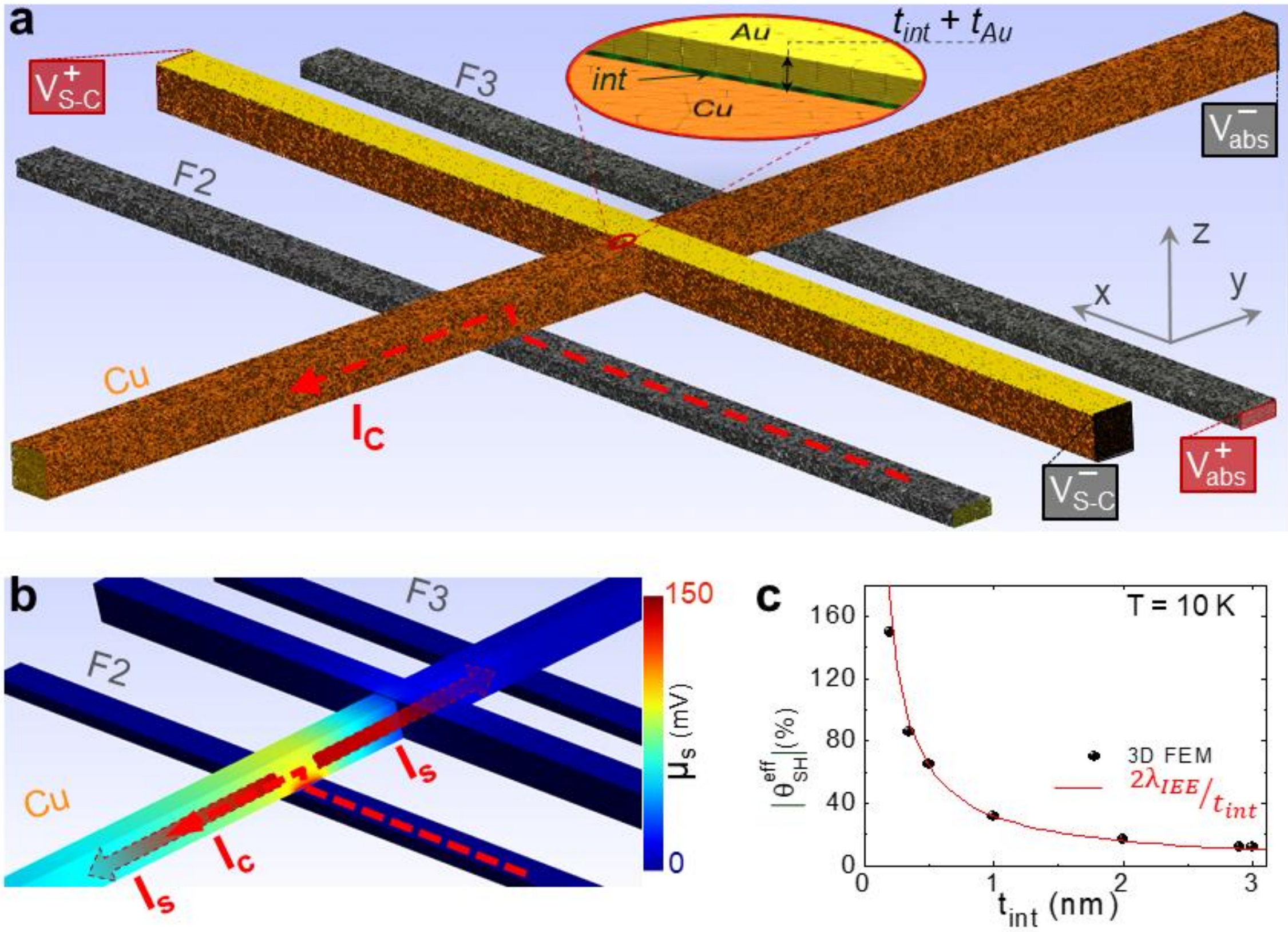

**FIG.9: 3D finite elements method simulation**. **a)** Geometry and mesh of the 3D FEM model used for simulating the spin absorption and spin-to-charge conversion signals using the two-current drift-diffusion equations. Inset: zoom of the mesh presenting the method to simulate the interface. The Au wire is separated as two layers in *z* direction. The first layer is an effective interface with spin-orbit coupling (*int*, in blue) with the resistivity of Au, thickness $t_{int}$, and two free parameters $\lambda_{int}$ and $\theta_{SH}^{eff}$. The second layer is the rest of Au with the inputs of bulk Au parameters ($\lambda_b^{Au}$and $\theta_{SH}^{Au}$) based on its resistivity (see Refs. 61 and 36). A current ($I_C$) is applied from F3 to Cu to induce a spin accumulation. **b**) Representation of the spin accumulation ($\mu_s$) landscape, i.e., the difference of electrochemical potentials given by up-spin and down-spin electrons. A pure spin current ($I_s$) is injected in the Cu channel in *y* direction. This $I_s$ is relaxed and converted to a charge current due to spin Hall effect in both the interfacial and the Au layers. The SC signal amplitude is the difference between to electrical potentials at the ends of the wire and normalized to the applied current [$\Delta R_{SC} = (V_{SC}^{+} - V_{SC}^{-})/I_C$]. This corresponds to the experimental results in Fig. 3b. The rest of $I_s$ induces a spin accumulation at F3/Cu interface which is probed in the same way as [$\Delta R_{NL}^{abs} = (V_{abs}^{+} - V_{abs}^{-})/I_C$]. This value is fit with the spin absorption signal to extract $\lambda_{int}$. Since all equations used for the simulations are linear, $I_c$ has been set to 1 A. **c**) Black spheres are the effective spin Hall angle $\theta_{SH}^{eff}$ obtained for each thickness $t_{int}$ in the FEM model to match the experiment results. The red curve is the fit to Eq. (3) of the main text.

First, we verify the spin transport parameters of Cu and NiFe by performing the simulations for a nonlocal measurement in a LSV. The spin diffusion lengths of NiFe and Cu are extracted from Ref. [64], by taking into account that their spin resistances ($\lambda_{NiFe}\rho_{NiFe} = 0.9$ f$\Omega$m$^2$ and $\lambda_{Cu}\rho_{Cu} = 2.3$ f$\Omega$m$^2$) have to be constant. The spin polarization of NiFe is taken from Ref. 64. The simulated spin signal reproduces well the value of the reference spin signal ($2\Delta R_{NL}^{ref}$) shown in Fig. 1b of the main text. Hence, we use these parameters to simulate the spin absorption and spin-charge interconversion.

To simulate the spin absorption and the SC conversion at the Au/Cu interface, a thin layer ($t_{int}$) is included between the Au and Cu layers in the vertical channel, as shown in the inset of FIG.9a. The thickness of the bulk Au ($t_{Au}$) is chosen so that $t_{int} + t_{Au}$ = 3 nm. The spin diffusion length ($\lambda_{int}$) accounts for the spin absorption at the interface and an effective spin Hall angle ($\theta_{SH}^{eff}$) accounts for the SC conversion at the interface. The spin diffusion length ($\lambda_b^{Au}$) and the spin Hall angle ($\theta_{SH}^{Au}$) of the top bulk Au layer are chosen based on the measured $\rho_{Au}$ (Appendix 2). $\lambda_b^{Au}$ is obtained from the spin resistance of bulk Au $\lambda_b^{Au}\rho_{Au} = 4.5$ fΩm$^2$ (Ref. 61), whereas $\theta_{SH}^{Au}$ is obtained based on the intrinsic contribution to SHE in bulk Au, *i.e.,* intrinsic spin Hall conductivity of 360 (ħ/e)$\Omega^{-1}$cm$^{-1}$ or $\lambda_b^{Au}\theta_{SH}^{Au} = 0.09$ nm (see Ref. 36).

**Table 1:** The parameters and results of the 3D FEM simulations at 10 K. $t_{int}$ is the defined effective thickness of the interface. $\lambda_{int}$ is the retrieved spin diffusion length by modeling the spin absorption experiment. $\theta_{int}^{eff}$ is the effective spin Hall angle associated to the interface by modeling the experimental $2\Delta R_{SC}$ values. $\Delta R_{SHE}^{Au}$ is the contribution of the SHE by the rest of bulk Au which is not considered interface. $\lambda_{IEE}$, $G_{\parallel}$, and $\sigma_{SC}$ are the inverse Edelstein length, interfacial spin-loss conductance, and spin-to-charge conductivity, respectively, calculated extracted using the relations between the FEM and the boundary conditions models (see text).

| $t_{int}$ (nm) | $\lambda_{int}$ (nm) | $\theta_{int}^{eff}$ | $\Delta R_{SHE}^{Au}$ (μΩ) | $\lambda_{IEE}$ (nm) | $G_{\parallel}$ ($\Omega^{-1}$m$^{-2}$) | $\sigma_{SC}$ ($\Omega^{-1}$cm$^{-1}$) |
|---|---|---|---|---|---|---|
| 3 | 32.5 | -0.11 | 0 | -0.17 | $7.6\times10^{13}$ | -126 |
| 2 | 25.8 | -0.17 | 0.5 | -0.17 | $8.0\times10^{13}$ | -132 |
| 1 | 18,3 | -0.33 | 0.7 | -0.17 | $8.0\times10^{13}$ | -131 |
| 0.5 | 13.2 | -0.67 | 1.0 | -0.17 | $7.7\times10^{13}$ | -128 |
| 0.35 | 10.6 | -0.86 | 1.2 | -0.15 | $8.3\times10^{13}$ | -125 |
| 0.2 | 8.3 | -1.50 | 1.3 | -0.15 | $7.7\times10^{13}$ | -116 |

The spin accumulation profile is illustrated in FIG.9b when applying a charge current from the electrode F2 to the Cu channel, which shows the spin current is absorbed by the effective interface and the Au wire. We adjust $\lambda_{int}$ in the FEM simulation to reproduce the experimental $\Delta R_{NL}^{abs}$ values. The obtained $\lambda_{int}$ is shown in Table 1. By using the obtained values of $\lambda_{int}$ and adjusting $\theta_{SH}^{eff}$ in the simulation in order to reproduce the experimental $2\Delta R_{SC}$ values shown in Fig. 3b in the main text, we obtain $\theta_{int}^{eff}$. Reciprocally, the same value of CS signal is achieved when using the obtained $\theta_{int}^{eff}$ and $\lambda_{int}$ in the CS conversion model. The simulations are repeated for different thicknesses between 0.2 and 3 nm of the effective interfacial layer, yielding different values of $\theta_{SH}^{eff}$ and $\lambda_{int}$ which always show $\lambda_{int} \gg t_{int}$. Figure S7c plots $\theta_{SH}^{eff}$ as a function of $t_{int}$ at 10 K. By fitting it with Eq. (3) of the main text, we achieve a constant $\lambda_{IEE} = -0.17 \pm 0.01$ nm at 10 K. Note that the contribution of SHE in the bulk Au ($\Delta R_{SHE}^{Au}$) is opposite in sign to the interfacial effect but very small. The 3D simulation shows that the maximum value of $\Delta R_{SHE}^{Au}$ at 10 K, which occurs with the smallest $t_{int}$ chosen, contributes only –2.3% to the total SC signal (see Table 1). Hence, for the other temperatures, we performed the simulations with $t_{int}$ = 3 nm and did not consider any contribution from the SHE of bulk Au. The effective spin Hall angle at different temperatures is plotted in Fig. 3c of the main text and the effective spin diffusion length $\lambda_{int}$ at different temperatures is illustrated in FIG.10.

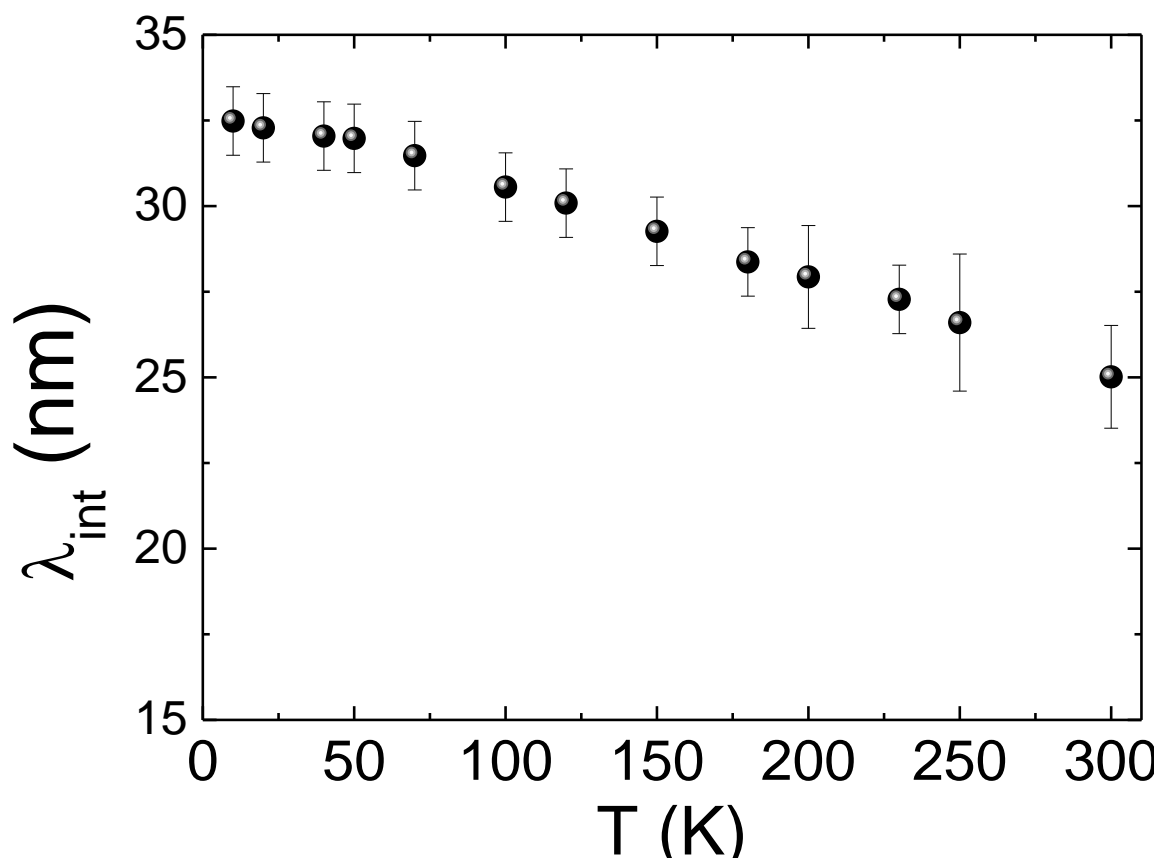

**FIG.10:** Effective spin diffusion length $\lambda_{int}$ as a function of temperature. The values are extracted by modelling the spin absorption results with the FEM simulation using $t_{int} = 3$ nm.

Finally, we would like to emphasize that the effective spin Hall angle $\theta_{SH}^{eff}$ is strongly dependent on the choice of the effective thickness $t_{int}$ for the interface, so that we can obtain any value of $\theta_{SH}^{eff}$ associated to such interface. However, the obtained $\lambda_{IEE}$, $G_{\parallel}$, and $\sigma_{SC}$ are independent of $t_{int}$, confirming they are the proper parameters to quantify the SC conversion at an interface.